\documentclass[twocolumn,showpacs,preprintnumbers]{revtex4}
\usepackage{graphicx}

\begin{document}

\title{Comparison of Monte-Carlo and Einstein methods in the light-gas interactions.}

\author{Jacques Moret-Bailly}
\email{jacques.moret-bailly@laposte.net}
\date{\today}

\begin{abstract}
To study the propagation of light in nebulae, many astrophysicists use a Monte-Carlo computation which does not take interferences into account. Replacing the wrong method by Einstein coefficients theory gives, on an example, a theoretical spectrum much closer to the observed one.
\end{abstract}
\maketitle

Transposition into optics of the Monte Carlo computations used in nuclear physics is wrong because it does not take into account the phases of the pilot-waves of the photons. Indeed, this approach denies the existence of interferences because the interaction of a neutron with an atom of uranium is multiform, complex and often slow, so that the phase of the incident neutron is lost. On the contrary, for example, when two photons of same frequency and same polarization simultaneously strike an atom, whether their phases are same or opposite is important.

\begin{figure}[l]{}
   \includegraphics[width=10cm]{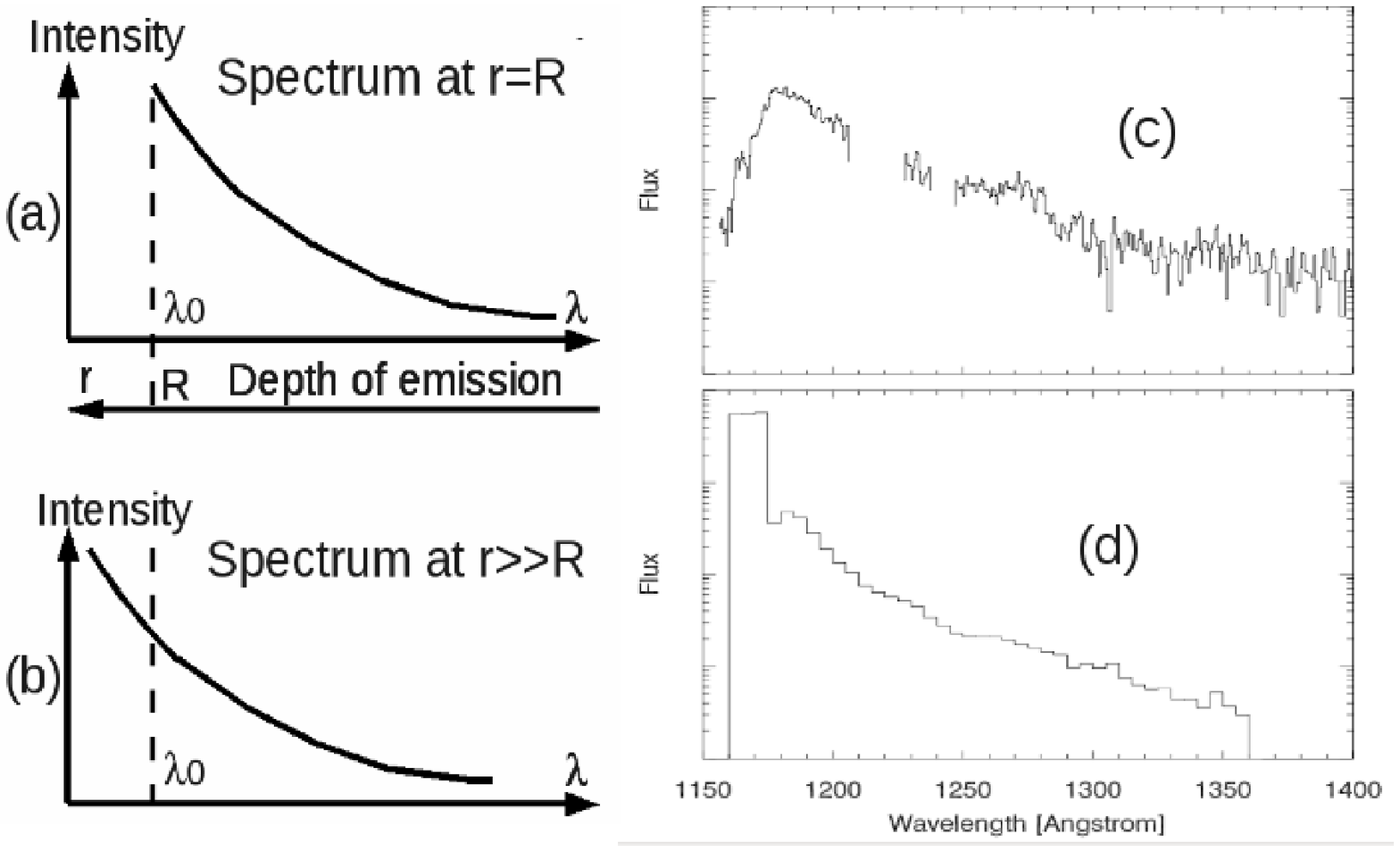}
   \caption{Spectra of Lyman $\alpha$ line inside the main ring of SNR1987A; a and b present theory; c experimental and d theoretical, from \cite{Michael}}
\end{figure}

\medskip
The correct method for studying the propagation of light in a resonant medium has been introduced by Einstein \cite{Einstein1916}, but it remains disputed in astrophysics, as it was by most physicists when Townes built the first maser. A sentence of Menzel written in 1931 \cite{Menzel} continues to be applied: \textquotedblleft It is easily proved that the so-called \textquotedblleft stimulated emissions\textquotedblright{} are unimportant in the nebulae\textquotedblright. Menzel got this result confusing the radiance of a ray, independent on distance from the source in a transparent medium, with irradiance, which decreases with distance. Some astrophysicists, for example the 21 authors of the article by Michael et al. \cite{Michael} use the Monte-Carlo method without considering the phase of the pilot-wave of the photons. 

The method of Einstein coefficients is simpler and can be used in two equivalent ways, depending on whether one uses the absolute radiance of light deduced from the papers of Planck \cite{Planck1911}, Einstein and Stern \cite{Einstein1913}, or the usual, relative radiance posed null beyond an opaque, cold screen. For a progressive mode, that is a beam of monochromatic light of frequency $\nu$ limited by diffraction, the relative radiance is deduced from the absolute radiance by adding $h\nu^3/c^2$. Using the absolute radiance, coefficient A is zero and the spontaneous emission which corresponds to the amplification of the minimal radiance $h\nu^3/c^2$ is considered as coherent because the amplification preserves the phase of the minimal, stochastic field.

Based on thermodynamics, statistical theory, the Einstein method is valid only if volumes of homogeneous material are large enough for the application of the law of large numbers, but there is no other limitation. Otherwise, fluctuations of density produce an opalescence, incoherent interaction of light with matter. At low pressure, these fluctuations result from collisions. The number of binary collisions in a given volume, at low pressure, is proportional to the square of pressure, so that incoherent interactions are negligible in astrophysics, except, of course, near the stars.

The simplest coherent emission is Rayleigh coherent scattering, at the incident frequency. Coherence allows for simple addition of the amplitudes of scattered field and incident field. By adding first the fraction of amplitude scattered in quadrature, refraction is obtained. If it is linear, the modes are preserved. The fraction of amplitude scattered in phase (or antiphase) produces amplification (or absorption).
The scattering of a different frequency produces waves of generally different wavelength, so that the scattered amplitudes add with different phases, self destroying. Therefore the final contribution of scattering is generally negligible. The use of lasers and crystals, for example, can however produce intense effects. G. L. Lamb \cite{Lamb} describes the use of pulses \textquotedblleft shorter than all relevant time constants\textquotedblright. With these conditions, the linewidths are larger than the distance between the exciting and a Raman lines which are mixed. An elementary computation shows that the scattered wave interferes with the exciting wave to form a sine wave of slightly different frequency, and negligible residual self-destructive waves. Thus, the frequency of laser pulses that transmit information over long distances through optical fibers are lowered enough to cause problems for color selection. These redshifts come from the excitation of quadrupoles whose energy is transferred simultaneously to thermal radiation by similar scatterings, the whole forming a parametric interaction.

The time constants considered by Lamb are, in a gas, the collisional time and the period of the quadrupolar transition. Using the pulses forming natural time-incoherent light, long of about a nanosecond, a dilute gas and a frequency below 1 GHz must be used. For example, the frequencies 178 MHz in the 2S$_{1/2}$ state of atomic hydrogen, 59 MHz in 2P$_{1/2}$ state, and 24 MHz in 2P$_{3/2}$ are very convenient while 1420 MHz is not. The effect is very weak, but can accumulate along an astronomical distance.

The spectrum of SNR1987A that Michael et al. studied results of the generation and propagation of the Lyman alpha hydrogen line in the outer region of a sphere of radius R, made of hydrogen almost fully ionized by its high temperature. An increase of the distance $r$ from the center of the sphere increases the density of neutral atoms, almost exponentially. The increase of $r$ decreases the path in excited atomic hydrogen across to radius R, along which the frequency is decreased. This gives Figure 1a. The global increase of frequency showed in Figure 1b results from the path of light in a region outside the sphere where beams of high radiance similar to those seen in the form of ring transfer energy to the spontaneously emitted light.

\medskip
Conclusion. 
Michael et al. seek an explanation for the large redshift in a spectrum of SNR 1987A. The introduction of coherence corrects their work, cutting their computed huge peak intensity. The introduction of coherence strengthens their search for an alternative to the standard explanation of large redshifts, so to the Hubble law, putting into question the basis of the theory of the big bang.

\end{document}